\documentclass[11pt,a4paper]{article}
\usepackage{jinstpub}

\usepackage{graphicx}
\usepackage{amssymb}
\usepackage{amsmath}
\usepackage{amsfonts}
\usepackage{dcolumn}

\def\Acal{{\cal A}}

\def\bv{{\bf b}}

\def\gv{{\bf g}}

\def\jv{{\bf j}}
\def\pv{{\bf p}}
\def\qv{{\bf q}}
\def\rv{{\bf r}}
\def\sv{{\bf s}}
\def\uv{{\bf u}}

\def\Bv{{\bf B}}

\def\Ev{{\bf E}}

 \def\Nv{{N}} % \def\Nv{{\boldsymbol{\nu}}}

\def\Rv{{\bf R}}

\def\eT{E_{\perp}}
\def\Lin{^{\rm (Lin)}}
\def\per{^{\rm (per)}}
\def\vib{^{\rm (vib)}}

\def\el{^{\rm (e)}}
\def\atom{_{\rm A}}
\def\cl{{\rm cl}}

\def\seclas{\hat\sigma^{\rm (cl)}}
\def\secqu{\hat\sigma^{\rm (quant)}}

\def\be{\begin{equation}}
\def\ee{\end{equation}}
\def\ni{\noindent}

\def\lambdabar{\lambda\raise0.4ex\hbox{\kern-0.5em\hbox{--}}\ }
\def\lambdaC{\lambda\raise0.5ex\hbox{\kern-0.5em\hbox{--}}_{\rm C}}
\def\lambdabarc{\lambda\raise0.5ex\hbox{\kern-0.5em\hbox{--}}_{\rm c}}
\def\lesssim{\,{\lower0.5ex\hbox{$\stackrel{<}{\sim}$}}\,}
\def\gtrsim{\,{\lower0.5ex\hbox{$\stackrel{>}{\sim}$}}\,}

\def\T{_{\rm T}}   

\def\pp{_\perp}
\def\pl{_\parallel}
\def\A{_{\rm A}}

\title{\boldmath Quantum versus classical approach of dechanneling and incoherent electromagnetic processes in aligned crystals (draft)}

\author{X. Artru} 
\affiliation{Universit\'e de Lyon, Institut de Physique des deux Infinis (IP2I Lyon),  \\
Universit\'e Lyon~1 and CNRS, France}
\emailAdd{x.artru@ipnl.in2p3.fr}

\abstract{
Particles traveling in aligned crystals at small angles w.r.t. crystallographic axes or planes 
are principally steered by the continuous Lindhard potential. This interaction conserves the energy E, the longitudinal momentum $p\pl$, the transverse energy of the particle $E\pp$ and is elastic concerning the crystal quantum state. At high enough energy the particle motion is quasi-classical. The time-dependent fluctuations of the positions of the atoms or of the electrons of the crystal create a residual potential, on which the particle can scatter. This interaction does not conserve the previous quantities and is inelastic for the crystal.
We compare its treatments with the classical binary collision model and with a phenomenological quantum model.
The classical dechanneling rate is estimated to be several ten per cent larger than the quantum one. 
The influence of correlated vibrations of neighboring atoms is discussed. }

\begin{document}
\maketitle % ici pour RevTeX
\flushbottom

%\begin{keyword}
%channeling \sep dechanneling \sep phonons \sep coherent bremsstrahlung
%% \MSC[2010] ...\sep  ...
%\end{keyword}

\section{Introduction}

A fast charged particle traveling in a crystal at small angle w.r.t. crystallographic axes or planes 
can be channeled by the continuous \textit{Lindhard electric potential}, $V\Lin(\rv\pp)$, where
$\rv\pp=(x,y)$ for axial channeling, $\rv\pp=x$ for planar channeling.
This potential is obtained from the time-dependent microscopic potential $V(t,\rv)$ by averaging over time and over the parallel coordinate(s), $\rv\pl=z$ for axial channeling, $\rv\pl=(y,z)$ for planar channeling. 

% [It is smooth at the scales below the lattice vibration amplitude  $u_1\sim 10^{-2}$ nm.]

 We consider particles of charge $\pm e$. The interaction $\pm e V\Lin(\rv\pp)$ conserves the total energy $E$, the parallel momentum $\pv\pl$  and the transverse energy of the particle, $\eT\simeq \pv\pp^2/(2E) \pm e V\Lin(\rv\pp)$.%
\footnote{We use the natural units system $c=\hbar=1$ $\,\alpha=e^2/(4\pi)\simeq1/137$.}
It is also responsible for channeling radiation and coherent bremsstrahlung. In computer simulations for high enough energies ($E\gtrsim100$ MeV for electrons), the particle motion steered by $V\Lin$ is described in terms of smooth classical trajectories. 
The {\it residual potential}, 
\be \label{residual} 
\delta V(t,\rv) = V(t,\rv) - V\Lin(\rv\pp) \,,
\ee
is a perturbation with respect to the channeling Hamiltonian $H_{\rm ch} = (\pv^2\!+\!m^2)^{1/2} \pm e V\Lin(\rv\pp)$.
It conserves none of the quantities $E$,  $\pv\pl$ and $\eT$.  It causes dechanneling, volume capture and incoherent bremsstrahlung. It  decomposes in \cite{Daba} 
\begin{eqnarray} \label{perturb} % (\ref{perturb})
\delta V(t,\rv)=  \delta V\per(\rv) + \delta V\vib(t,\rv) + \delta V\el(t,\rv) \,.
\end{eqnarray}
Each term of  Eq. (\ref{perturb}) corresponds to a specific time scale, degree of inelasticity and collective motion of the crystal constituents. 
\begin{description}
\item[$\bullet$] 
$\delta V\per$ takes the periodic structure of atomic strings or planes into account. $V\Lin + \delta V\per$ is the time-average of $V(t,\rv)$. It conserves $E$ but not $\pv\pl$ and $\eT$.
\item[$\bullet$] 
$\delta V\vib(t,\rv)$ is due to the lattice vibrations (phonons + zero-point motion).  
Its shortest time scale is $\Delta t\vib\sim\max \{\omega_{\rm phonon}\}$. 
Scattering by $\delta V\vib$ creates or annihilates phonons and may dislodge atoms from their sites. 
It is inelastic for the crystal but elastic for the atoms.
\item[$\bullet$] 
$\delta V\el(t,\rv)$ represents the fluctuations due to the electron orbital motions. Its time scale is $\Delta t\el \sim \hbar$ (eV)$^{-1}\ll\Delta t\vib$. Scattering by $\delta V\el$ excites or ionizes the atoms. 
\end{description}
The charge density $\rho(t,\rv)=  - \nabla^2V(t,\rv)$ is decomposed in the same way:
\begin{eqnarray}
\rho(t,\rv) = \rho\Lin(\rv\pp) + \delta \rho\per(\rv) + \delta \rho\vib(t,\rv) + \delta \rho\el(t,\rv) \,.
\end{eqnarray}
$\delta V\vib$ and $\delta V\el$  are the sums of contributions  $\delta\hat V_{\Nv}\vib$ from individual atoms and $\delta\hat V_{\Nv,k}\el$ from individual electrons. We call such contributions ``thorns'', because they have a $1/r$ Coulomb peak at the instantaneous position of the nucleus or electron. 
%  (for $\delta\hat V_{\Nv}\vib$) (for $\delta\hat V_{\Nv,k}\el$). 

In this paper we will compare two different approaches for treating the perturbation $\delta V$ % (\ref{perturb})
 in computer simulations: 
\begin{description}
\item[ 1)] the purely classical model (CM), where the trajectory is calculated with the classical equation of motion in the full instantaneous potential $V(t,\rv)=V\Lin(\rv\pp)+\delta V(t,\rv)$.
\item[ 2)] a semi-classical model (SCM), where the trajectory is calculated classically in the potential $V\Lin$, but redirected from place to place by a scattering on a `thorn' $\delta V_{\Nv}\vib$ or $\delta V_{\Nv,k}\el$. The scattering angle is randomly chosen according to a quantum-mechanical differential cross section. A more developed approach based on Wigner distributions is found in Ref. \cite{Tikho19}. 
\end{description} 
Both approaches use the Monte Carlo method. In CM, the random variable is the instantaneous position of the incoherent scatterer (atom or electron). In SCM,  the random variable is the momentum transfer $\qv$. 

In Section 2 we present the characteristic shapes of the atomic and electronic `thorns'. Section 3 is devoted to the comparison between the classical and quantum differential cross sections on a `thorn'.
The effects of correlations between the vibrations of neighboring atoms about their equilibrium positions is discussed in Section 4.

\section{\boldmath Peculiarities of $\delta V\per$, $\delta V\vib$ and $\delta V\el$}

\paragraph{{\boldmath The potential $\delta V\per(\rv)$}$.$}

It is smooth on a scale smaller than the lattice vibration amplitude $u_1\sim 10^{-2}$ nm. Scattering on it is 
a Bragg diffraction involving only the $\gv\pl\ne0$ reciprocal lattice vectors, since the $\gv\pl=0$ ones are in $V\Lin$.
We have $\gv^2 = 2 |\gv\cdot\pv|$ from conservation of $E$. In axial channeling $\gv^2 \simeq 2E |g\pl| \ge 4\pi E/$({\it lattice constant}) is large, therefore the scattering is suppressed by the Debye-Waller factor $\exp\!\left(-\gv^2 u_1^2\right)$. A more involved reasoning leads to the same conclusion for planar channeling, if far from a major axis. From now on we will neglect $\delta V\per$.

\paragraph{{\boldmath The potential $\delta V\vib(t,\rv)$}$.$}
Its decomposition in `thorns' from individual atoms is $\delta V\vib(t,\rv) = \sum_\Nv \delta \hat V\vib_\Nv(t,\rv)$, where $\Nv=\{n_1,n_2,n_3\}$ labels the crystal lattice site and 
\be   \label{dVA} % (\ref{dVA})
 \delta \hat V\vib_\Nv(t,\rv) = V\atom(\rv-\Rv_\Nv(t)) - \overline V\atom(\rv-\overline\Rv_\Nv) \,;
\ee
%F
 $\overline{\Rv}_\Nv$ and $\Rv_\Nv(t)=\overline{\Rv}_\Nv + \uv_\Nv(t)$ are the mean and instantaneous positions of the nucleus.  $V\atom(\rv)$ is the atomic potential and $\overline V\atom=V\atom\otimes D$ is its convolution with the 
distribution of the atom displacement, $D(\uv)= (2\pi u_1^2)^{-3/2} \exp[-\uv^2/(2u_1^2)]$.
The atomic `thorn', also referred to as `remnant atom' in literature, produces less scattering than an atom in amorphous matter \cite{Kagan,Daba} (relevant references are also found in Ref. \cite{Tikho19}).
Equation (\ref{dVA}) corresponds to Eq. 4 of Ref. \cite{Tikho19}. 
$V\atom$ possesses the singularity $(4\pi)^{-1}Ze/|\rv \!-\! \Rv_\Nv(t)|$ coming from the Coulomb potential of the nucleus. 
The potential $\overline V\atom(\rv)$ is smooth at scales $<u_1$. 
$\delta \hat V\vib_\Nv$ is not spherically symmetric and its integral  
$
\int  \! d^3\rv \, \delta \hat V\vib_\Nv(\rv) 
$
over space is zero. Most of the times, $|\uv_\Nv| \sim u_1$ 
is much smaller than the Thomas-Fermi radius $a_{\rm TF}$, 
therefore $V\atom$ and $\overline V\atom$ almost cancel in the intermediate range $[u_1,a_{\rm TF}]$, making $\delta \hat V\vib$ weak and of very short range.  
Figure 1 (a) shows a typical transverse profile of $\delta \hat V_\Nv\vib(t,\rv)$.

%
%  FIGURE  FIGURE  FIGURE  FIGURE  FIGURE  FIGURE  FIGURE  FIGURE  
\begin{figure}   % [b]  % =Fig.1
 \centering
\includegraphics*[angle=90, width=0.8\textwidth, bb= 180 80 435 745]
{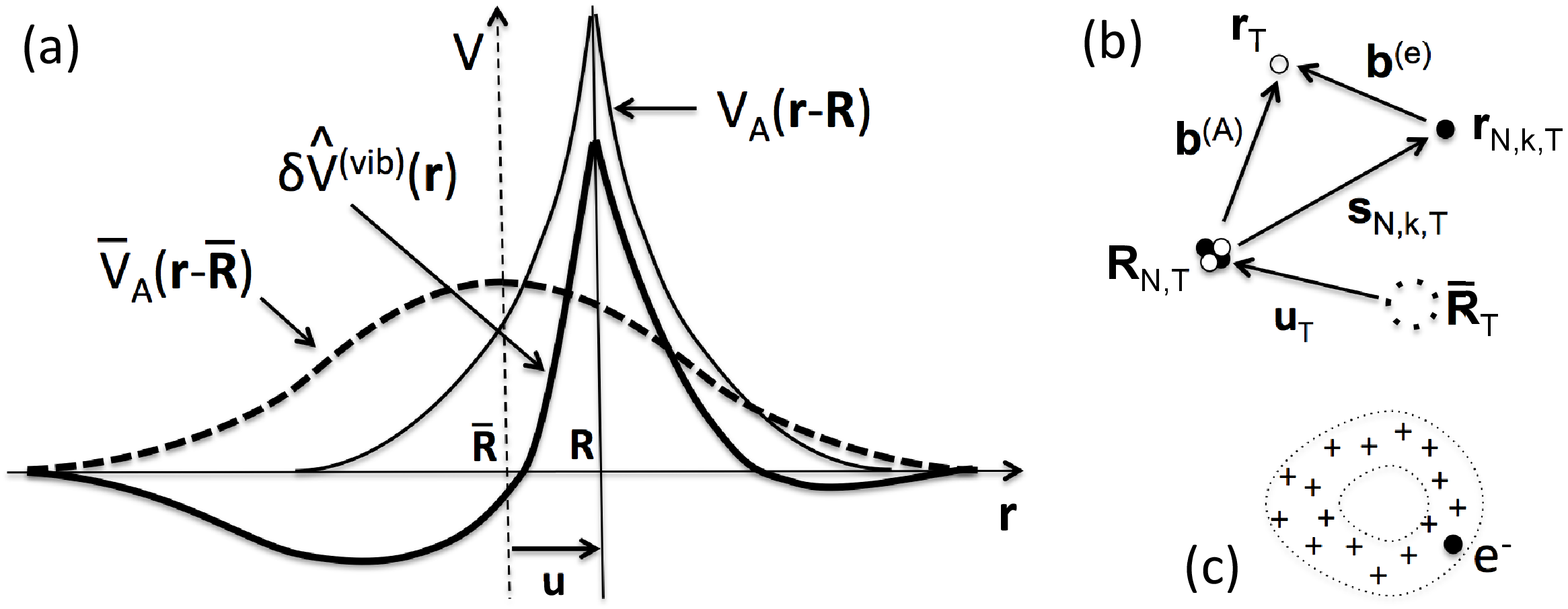}
\caption{%\footnotesize  
(a) Profiles of $V\A$, $\overline{V}\A$ and $\delta \hat V_\Nv\vib(\rv)$ (Eq. \ref{dVA}). To prevent the peak at $\rv=\Rv$ from being infinitely high, the potentials are integrated over the dimensions perpendicular to the displacement $\uv$ of the atom.
(b) Representations of the vectors $\uv\T$, $\sv_{\Nv,k,{\rm T}}$ perpendicular to the particle trajectory, and of the impact parameters $\bv^{(A)}$ and $\bv^{(e)}$. The white bullet is the particle, the black one is an atomic electron. 
% $\Rv_{\Nv}$, $\Rv_{\Nv,k}$,  $\Rv_{\Nv}$, $\Rv_{\Nv,k}$     
(c) Slice of the charge distribution $\delta \hat\rho_{\Nv,k}\el(\rv)$  (Eq. \ref{dVe}). A positively charged cloud,  represented by the annulus, takes the place of the electron orbital.
}
\end{figure}

\paragraph{{\boldmath The potential $\delta V\el(t,\rv)$}$.$} 
 Let us first consider the associated {\it charge density} $\delta \rho\el(t,\rv)$. It is a double sum $ \sum_\Nv \sum_{k=1}^{Z}\delta \hat\rho\el_{\Nv,k}$ over the atom site $\Nv$ and over the $Z$ electrons of the atom, with
\be   \label{dVe} % (\ref{dVe})   
\delta \hat\rho\el_{\Nv,k}(t,\rv) =  -e \left[\delta(\rv-\rv_{\Nv,k}(t)) - |\psi_{k}(\rv-\Rv_{\Nv}(t)) |^2
\right] ;
\ee
$\rv_{\Nv,k}(t) =\Rv_{\Nv}(t)+\sv_{\Nv,k}(t)$ 
is the instantaneous position of the $k^{\rm th}$ electron of the atom $\Nv$ [see Fig. 1 (b)] and $\psi_{k}(\rv)$ is its orbital wave function.
% neglecting electron correlations and the deformations of orbitals by chemical bounds.  
 The interpretation of the two terms in the square bracket in Eq. (\ref{dVe}) is the following:
a scattering which is inelastic for the atom $\Nv$ takes place at a definite time $t$. At this time, the electron is at $\rv_{\Nv,k}(t)$, yielding the delta function, represented in Fig. 1 (c) by a bullet. 
Being at $\rv_{\Nv,k}(t)$, the electron is absent in the other points of the orbital, producing there a defect of electronic density, represented in Fig. 1 (c) by the diffuse shell. This  gives the second term.   

 $\delta V\el(t,\rv)$ decomposes like $\delta \rho\el(t,\rv)$. From Eq. (\ref{dVe}) one obtains the contribution (`thorn') of the electron numbered $\Nv,k$: 
\be \label{dVek} % (\ref{dVek}) 
\delta \hat V\el_{\Nv,k}(t,\rv) =  -(4\pi)^{-1}   \left[e/|\rv-\rv_{\Nv,k}(t)| - |\psi_{k}(\rv-\Rv_{\Nv}(t)) |^2 
 \otimes e/|\rv| \right].
\ee
The first term in the square bracket of Eq. (\ref{dVek}) has the Coulomb singularity at the instantaneous electron location. The second term yields a regular potential with a tail in $1/r$ canceling that of the first term, resulting in a {\it dipolar} asymptotic behavior $-e\,[\sv_{\Nv,k}(t)-\langle\sv_{\Nv,k}\rangle]\cdot\rv/r^{3}$. 
{A figure analogous to Fig. 1(a) is obtained when plotting $\delta \hat V\el_{\Nv,k}$ and the two terms of the right-hand side of Eq. (\ref{dVek}).
}
% where $\sv_{\Nv,k}(t)\equiv\rv_{\Nv,k}(t)-\Rv_\Nv$ is the relative coordinate of the electron in the atom.  

\paragraph{{\boldmath Other perturbations to $H_{\rm ch}$}$.$} 
Associated to the fluctuating charge density $\delta\rho\el(t,\rv)$ is a current density $\jv\el(t,\rv)$, producing a microscopic magnetic field $\Bv\el(t,\rv)$. 
Since the crystal nuclei and electrons are non-relativistic, this field is much smaller than the fluctuation $\delta\Ev\el$ of the electric field, therefore neglected.
%  since it is much smaller than the electric filed $\delta\Ev\el$ created by the non-relativistic crystal constituents. 
%The nucleus position also fluctuates at the atomic time scale, whence a contribution 
%$\delta \hat V^{\rm (nucl)}_{\Nv}$ in addition to $\delta \hat V\el_{\Nv,k}$.
%However, the nucleus displacement is smaller than the electron ones by a factor $m_{\rm e}/m_{\rm nucl.}$, therefore we neglect it. 
% We have also not included the inelastic collisions with nuclei, which occur relatively seldom. 
{We have also omitted the scattering on nuclei which are inelastic for the atom (for a momentum transfer $|\qv|\gtrsim \alpha M_{\rm nucleus}$) or for the nucleus itself. Such collisions rarely occur within one dechanneling length.}

\section{The two methods for computing the trajectory} 

Before hand, we replace $\delta V\vib(t,\rv)$ and $\delta V\el(t,\rv)$ by ``frozen'' potentials $\delta V\vib(\rv)$ and $\delta V\el(\rv)$, assuming that, before and after scattering, the crystal electrons and nuclei are very slow compared to the relativistic projectile. This is true except in the rare collisions on electrons at momentum transfer $|\qv|\gtrsim m\el$. Let us present in more details the two approaches mentioned in the introduction.

\paragraph{The classical model (CM).}

The particle trajectory is calculated using the classical equation of motion $d\pv/dt=\mp e\nabla V(\rv)$, $d\rv/dt=\pv/E$, where $V(\rv)$ is the instantaneous potential $V(\rv)=V\Lin(\rv\pp)+\delta V(\rv)$ (the argument $t$ is now omitted). To save computation time, only the Coulomb fields of constituents at distance smaller than, say, one lattice constant from the trajectory are taken into account. This is the basis of the {\it `snapshot'} binary collision method used in Ref. \cite{Kostyuk}, but here the recoil of the scatterer is neglected.  

\paragraph{The semi-classical model (SCM).}

Pieces of trajectory are calculated using the classical equation of motion in the Lindhard potential $V\Lin(\rv\pp)$. On top of this regular motion, scatterings by `thorns' 
% potentials $\delta \hat V\vib_{\Nv}$ or $\delta \hat V\el_{\Nv,k}$ 
occur from place to place. They are represented by kinks of the trajectory, corresponding to sudden changes of momentum $\qv=\pv'$ - $\pv$. Using the Monte Carlo method, the kinks are generated randomly at a rate proportional to the local density of scatterers (nuclei for $\delta \hat V\vib$, electrons for $\delta\hat  V\el$), weighted by the cross sections $\hat\sigma$ of their $\delta\hat V$. At each kink, $\qv$ is drawn randomly following the differential cross section $d\hat\sigma/d^2\qv$. $\hat\sigma$ and $d\hat\sigma/d^2\qv$ are calculated in quantum mechanics {\it as if the particle was in an eigenstate of $\pv$} (approximation $\Acal1$).

\paragraph{{\boldmath Quantum calculation of $d\hat\sigma/d^2\qv$}$.$}
Let $\delta\hat V= \delta \hat V\vib_{\Nv}$ or $\delta \hat V\el_{\Nv,k}$ be a generic `thorn' and $\delta\hat\rho=-\nabla^2\delta\hat V$ the corresponding charge density.  The weakness and short range of $\delta\hat V$ justifies the Born approximation, 
\be \label{Born} % (\ref{Born})
\frac{d\hat\sigma}{d^2\qv} =  \frac{\alpha}{\pi}  % T(\qv) \frac{\pm e}{2\pi} 
\left| \int \! d^3\rv \exp(-i\qv\cdot\rv) \, \delta\hat V(\rv) \right|^2
=   \frac{\alpha}{\pi|\qv|^4} % \frac{\pm e}{2\pi\qv^2}
\left| \int \! d^3\rv \exp(-i\qv\cdot\rv) \, \delta\hat \rho(\rv)  \right|^2 \,.
\ee

\ni {\it A) if the incoherent scatterer is an atom,} Eq. (\ref{dVA}) gives
\be  \label{secdifquA}    
d\hat\sigma/d^2\qv =  
 4\left[Z\alpha \,f\atom(\qv^2) /\qv^2\right]^2 \, \left|\exp(i\qv\cdot \uv) - \exp(-\qv^2 u_1^2/2) \right|^2 ,
\ee
where $f\atom(\qv^2) = \int \! d^3\rv  \, \exp(-i\qv \!\cdot\! \rv) \, \rho\atom(\rv)/(Ze)$ is the charge form factor of the atom ($f\atom\equiv 1$ for a bare point like nucleus). The index $\Nv$ of $\uv_\Nv$ is omitted.
The Debye-Waller factor $\exp(-\qv^2 u_1^2/2) $ is the ratio $\overline V\atom/V\atom$ in momentum space.  
Denoting by $\qv\T$ and $\uv\T$ the transverse parts of $\qv$ and $\uv$ relative to $\pv$, we have $\qv\simeq\qv\T$, therefore $d\hat\sigma/d^2\qv$ depends only on $\qv\T$ and $\uv\T$; it is asymmetrical. The total cross section writes:
\be
\hat\sigma\atom  (\uv\T) 
= 4 \pi Z^2\alpha^2 \int_0^\infty d\qv^2 \frac{f^2\atom(\qv^2)}{|\qv|^4}
\left[ 1 + \exp(-\qv^2 u_1^2) - 2 \exp(-\qv^2 u_1^2/2) J_0(|\qv| |\uv|) \right] ,
\ee
which depends only on $|\uv\T|$. 
We assume that {\it the impact parameter $\bv^{\rm (A)}   \equiv \rv\T-\Rv_{\Nv\rm T}$ of the particle w.r.t. the nucleus} [see Fig.1(b)] {\it is small compared to $u_1$} (approximation $\Acal2$), thus $\uv\T\simeq \rv\T-\overline\Rv_{\Nv\rm T}$. Then the probability $P\vib$ of an incoherent, but elastic, collision on the atom $\Nv$ is 
%       differential   /d^2\qv\T
\be \label{P_A} % (\ref{P_A})
P\vib = (2\pi u_1^2)^{-1} \,  \exp[-(\rv\T-\overline\Rv_{\Nv\rm T})^2/(2u_1^2)] \, 
\hat\sigma\atom (\uv\T = \rv\T-\overline\Rv_{\Nv\rm T} ) \,;
\ee
the factors in front of $\hat\sigma\atom$ give the nucleus probability density integrated along the particle trajectory.

\ni {\it B) if the incoherent scatterer is an electron,} Eq. (\ref{dVe}) gives
% and the second expression in (\ref{Born}) gives 
%
\be \label{secdifque} 
d\hat\sigma/d^2\qv =  %T(\qv) =  ({\mp 2\alpha}/{\qv^2}) \, \exp(-i\qv\cdot \Rv_{\Nv}) 
4 (\alpha/\qv^2)^2  \, \left| \exp(-i\qv\cdot \sv_{\Nv k})  - f_k(\qv) \right|^2 ,
\ee
where $f_k(\qv) = \int \! d^3\sv \exp(-i\qv \!\cdot\! \rv) \, |\psi_{k}(\sv) |^2$ is the form factor of the $k^{\rm th}$  orbital. The $f_k$ are constrained by the relation $f\atom(\qv^2) = 1- \sum_{k=1}^Z f_k(\qv)/Z$, derived from 
$\hat\rho\atom(\rv)= Ze \delta(\rv) - e \sum_k |\psi_{k}(\rv)|^2$. 
The differential cross section depends on $\qv\T$ and $\sv_{\Nv, k,\rm T}$ and the total one on $\sv_{\Nv, k,\rm T}$.
%For a given value of $|\sv_{\Nv,k}|$, 
%\be % CA NE MARCHE QUE POUR $\psi_{k}(\rv)$ A SYMMETRIE SPHERIQUE
%\hat\sigma_k(\sv_{\Nv,k}) = ... \pi \int d\qv^2 \frac{1)}{|\qv|^4}\left[ 1 + f_k^2(\qv^2) - 2 f_k(\qv^2) J_0(|\qv|.|\sv_{\Nv,k}|) \right] \ee
To get the analogue of Eq. (\ref{P_A}) we assume that {\it the impact parameter $\bv^{\rm (e)} \equiv \rv\T-\rv_{\Nv,k,\rm T}$ of the particle w.r.t. the atomic electron is small compared to the width of $\psi_{k}$} (approximation $\Acal2'$), thus $\sv_{\Nv,k,\rm T} \simeq \bv^{\rm (A)}$.
% Under this approximation, 
Then the probability of an inelastic collision with an atom of given instantaneous position $\Rv_\Nv$ writes
\be \label{P_el} % (\ref{P_el})
P\el(\bv^{\rm (A)})=  \hat\sigma_k(\sv\T=\bv^{\rm (A)})  \sum_k \int \! dz\, |\psi_{k}(\bv^{\rm (A)} ,z)|^2 \,.
\ee
{This expression depends on $\uv_N$ through $\bv^{\rm (A)} \equiv \rv\T- \overline\Rv_{\Nv\rm T} - \uv_{\Nv\rm T}$. If $\uv_N$ is not specified, it has to be convoluted with   
$(2\pi u_1^2)^{-1} \,  \exp[-\uv\T^2/(2u_1^2)]$, giving a probability depending only on $(\rv-\overline\Rv_{\Nv})\T$, like $P\vib$ in Eq.  (\ref{P_A}). }

Equations analogous to (\ref{P_A}) and (\ref{P_el}), not written here, give the differential probabilities $dP /d^2\qv\T$. These have an azimuthal asymmetry analogous to the one predicted in Ref. \cite{Tikho19}.

\section{Comparison between CM and SCM}

If one assumes that the density of incoherent scatters is roughly constant within the ranges of their `thorns', one can present the CM in a form similar to SCM, but replacing the quantum-mechanical cross sections   $d\secqu/d^2\qv$ by the classical ones,  $d\seclas/d^2\qv$. Under this approximation, the
comparison between the classical and quantum approaches reduces to the comparison between $d\secqu/d^2\qv$ and $d\seclas/d^2\qv$ on the `thorn' $\delta \hat V$. 
For computer simulations, due to the complexity of the `thorn' potentials which are asymmetrical, one replaces them phenomenologically by spherical potentials possessing the same Coulomb peak and equivalent short ranges, which are the two most relevant features. This is done, {\it e.g.}, in Ref. \cite{XAsimul}.

In the classical model, the momentum transfer is a function $\qv_\cl(\bv)$ of the impact parameter $\bv=\rv\T - \rv_{i\rm T}$, where $\rv_i$ is the position of the incoherent scatterer (atom or electron). In the high-energy, straight-line approximation,    
\be  \label{Cl}
\begin{aligned}
\qv_\cl(\bv) &= - \int_{-\infty}^\infty  dz \, \nabla_\bv \, \delta \hat V(\rv_{i{\rm T}}+\bv,z)
\\
\frac{d\seclas}{d^2\qv}  % \frac{d\sigma_\cl}{d^2\qv} 
&= \int d^2\bv \, \delta[\qv-\qv_\cl(\bv) ] \,.
\end{aligned}
\ee
Equations (\ref{Born}) and (\ref{Cl}) have the common properties, also valid for asymmetrical potentials:
% The ${d\sigma}/{d^2\qv}$ and ${d\sigma_\cl}/{d^2\qv}$ 
%
%\be
%\begin{aligned}
\begin{eqnarray}
\int \qv \, d^2\qv \, d\secqu/d^2\qv  &=& \int  \qv \,  d^2\qv \, d\seclas/d^2\qv = {\bf0}
\label{sumrule_q}
\\
\int  \qv^2 \,  d^2\qv \, d\secqu/d^2\qv &=& \int  \qv^2 \,  d^2\qv \, d\seclas/d^2\qv \equiv  \int   \qv^2 \,  d^2\bv \, \qv^2_\cl(\bv) 
\label{sumrule_q2}
\end{eqnarray}
%\end{aligned}
%\ee
%
Equation (\ref{sumrule_q}) insures that in average $\delta V$ produces no transverse force. Equation (\ref{sumrule_q2}), discussed in Ref. \cite{Bondarenco}, diverges for the potentials given by Eqs. (\ref{dVA}) and  (\ref{dVek}) if their Coulomb singularities are not regularized. 

\paragraph{Scattering on an atom.} 
For an atom, the Coulomb singularity of $\delta \hat V\vib$ is regularized by the finite size $r_{\rm N}$ of the nucleus. This cuts off $f\atom(\qv^2)$ in Eq. (\ref{Cl}) at $|\qv| \sim 1/r_{\rm N}$ 
and $\qv_\cl(\bv)$ in Eq. (\ref{secdifquA}) at $|\bv|\sim r_{\rm N}$, making Eq. (\ref{sumrule_q2}) convergent. Replacing  
$\delta \hat V\vib(\rv)$ by the phenomenological potential $(Z\alpha/r) \, [\exp(-r/r_{\rm max})-\exp(-r/r_{\rm N})]$,  with $r_{\rm max}\sim u_1$,  % $r_{\rm min}\sim r_{\rm N}$ 
the ratio of $d\secqu/d^2\qv$ and $d\seclas/d^2\qv$ to the pure Coulomb cross section are pictured in Fig. 2 \cite{Bondarenco-rainbow}. 
The dechanneling efficiency of an atomic `thorn' can be roughly measured by the % {\it dechanneling 
cross section, 
\be \label{sigmadech}
\hat\sigma_{\rm dech} = q_c^{-2}  \int_0^{q_c}  \qv^2 \,  d\qv^2 \, \frac{d\hat \sigma}{d\qv^2}
+  \int_{q_c}^\infty  \,  d\qv^2 \,  \frac{d\hat \sigma}{d\qv^2}
\simeq
4 \pi \left(\frac{Z\alpha}{q_c}\right)^2 \, [2\ln(q_c/q_{\rm min}) + 1]\,,
\ee
where $q_c = E\psi_c = \sqrt{2EU_0}$, $\psi_c$ is the critical Linhard angle, $U_0$ the channeling potential  depth; $q_{\rm min}\simeq 1/r_{\rm max}$ in the quantum model, $q_{\rm min}\simeq Z\alpha/r_{\rm max}$ in the classical model. We have assumed $q_c < q_{\rm max}\simeq 1/r_{\rm N}$ in the quantum model or $Z\alpha/r_{\rm N}$ in the classical model.
The second integral of  Eq. (\ref{sigmadech}) corresponds to dechanneling by a single collision. The first one represents a gradual increase of the transverse energy by multiple incoherent scattering at moderate momentum transfer. It is proportional to the area situated under the curve $q^4 d\sigma/dq^2$ and on the left of $q_c$ in Fig. 2.  
It shows that for $Z\alpha < 1$ we have $ \hat\sigma_{\rm dech}^\cl > \hat\sigma_{\rm dech}^{\rm quant}$.
As an example, for an electron of $E$ =1 GeV channeled in Silicon, $\psi_c\sim 1$ mrad, $q_c\sim$ 1 MeV, $r_{\rm max}=u_1\sim$ 0.075 {\AA} and equation (\ref{sigmadech}) gives $ \hat\sigma_{\rm dech}^\cl / \hat\sigma_{\rm dech}^{\rm quant}$ = 1.43. 
Considering the SCM model as more realistic than the CM model, one concludes that the classical model  overestimates the dechanneling rate. The same conclusion is drawn in Ref. \cite{Tikho19}.

 %
%  FIGURE  FIGURE  FIGURE  FIGURE  FIGURE  FIGURE  FIGURE  FIGURE  
\begin{figure} %  [b!]  % =Fig.2
 \centering
\includegraphics*[angle=90, width=0.9\textwidth, bb= 185 70 415 770]
{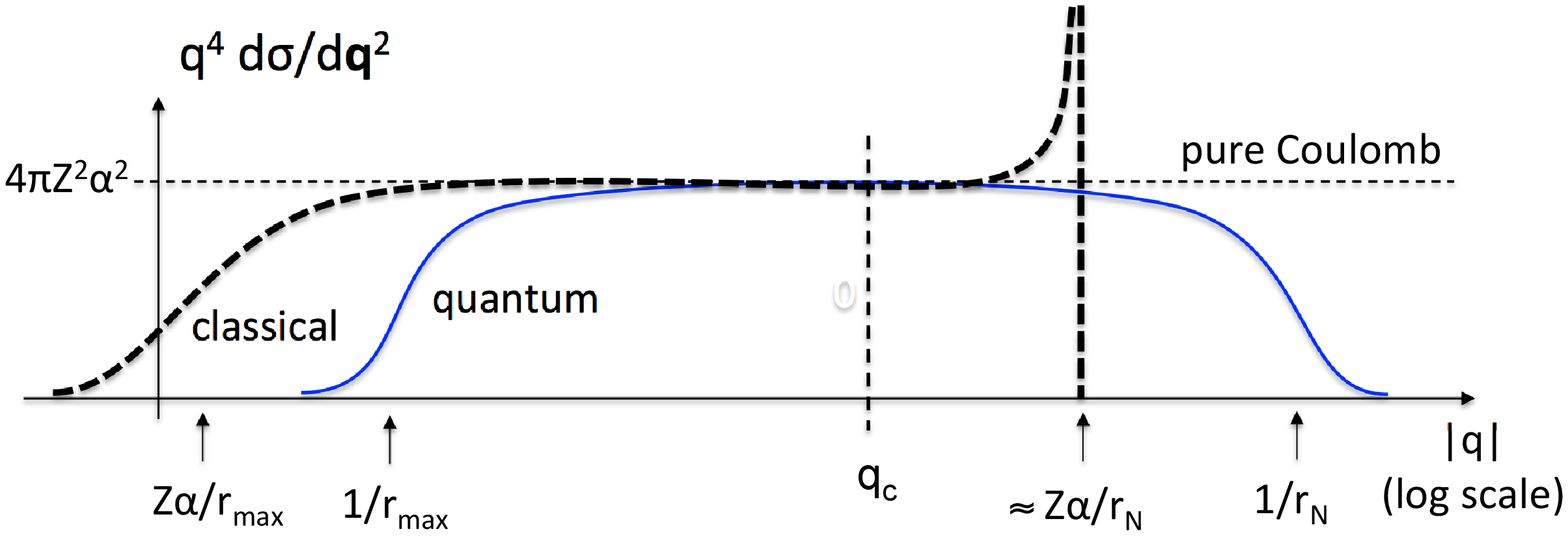}
\caption{%\footnotesize  
Comparison between the classical and quantum differential cross section for a Coulomb potential regularized at $r=r_{\rm min}\sim$ nuclear radius and screened at $r=r_{\rm max}\sim$ Thomas-Fermi radius. The classical cross section stops, with a rainbow peak, at $q\sim Z\alpha/r_{\rm N}$ due to the maximum of $|\qv_\cl(\bv)|$ \cite{Bondarenco-rainbow}. The area under the continuous curve (quantum case) and the dashed curve (classical case) are equal due to Eq. (\ref{sumrule_q2}). If $q>q_c$ dechanneling occurs at once.}
\end{figure}

\paragraph{Scattering on an electron.}
 The electron being point like, the cutoff $q_{\rm max}$ is very large and purely kinematical, corresponding to a scattering at $90^\circ$ in centre-of-mass frame. It is the same in the classical and quantum model, whereas their $q_{\rm min}$ differ by a factor $\sim$137, therefore Eq. (\ref{sumrule_q2}) is invalid. Equation (\ref{sigmadech}) is still valid, just replacing $Z$ by 1. Taking the same numerical example as above, but with $r_{\rm max}\sim a_{\rm TF}=$ 0.194{\AA}, one obtains $ \hat\sigma_{\rm dech}^\cl /  \hat\sigma_{\rm dech}^{\rm quant}$ = 1.97.   

 \section{Effect of the correlations between the $\uv_{N}$}
 
When considering the scattering by $\delta V\vib$ we have implicitly assumed that the displacements $u_\Nv$ of the atoms about their equilibrium positions are uncorrelated. 
In fact, this is not the case \cite{Glauber,Jackson-cor,Huan}. 
A long wave-length phonon moves many neighboring atoms in the same direction.
The effects of such correlations on the dechanneling length $L_{\rm d}$ for the axial case was investigated in Ref. \cite{XA-vibra} using the CM model. 
It was shown that they reduce $L_{\rm d}$. Indeed, the r.m.s. of the transverse distance $|\uv_\Nv-\uv_{N'}|$ between two neighboring atoms of a chain is much less than $\sqrt 2\, u_1$, so that these atoms are  likely to scatter the particle coherently, like one ``super-atom'' of atomic number $2Z$. In a SCM approach, one can take the effect of long wave-length phonons into account by redefining $\uv_\Nv$, $V\Lin$ and $\delta V\vib$. At fixed time, the new $V\Lin(\rv)$ follows the local smooth bends of the crystal generated by the long wave-length phonons, like in a crystal undulator. It is treated classically. $\delta V\vib$ is now due to short wave-length phonons only and treated quantum-mechanically, neglecting the correlations between the new $\uv_\Nv$. The new vibration amplitude $u_1$ is reduced accordingly, allowing stronger macroscopic fields. This should produce a {\it semi-coherent} bremsstrahlung \cite{XA-vibra} which adds to channeling radiation, {while the {\it true incoherent} bremsstrahlung is reduced}. Up to now, no quantitative study of this effect has been done. The separation between ``long'' and ``short'' wavelength phonons has to be defined.       

\section{Conclusion}

We have compared the classical (CM) and a semi-classical (SCM) models for treating incoherent scattering in channeling with Monte Carlo simulations. The CM, based on binary collisions, is theoretically clear. By comparison, the SCM, which needs the questionable approximations $\Acal1$, $\Acal2$ and $\Acal2'$,
is not so precise but more realistic.  Indeed, the classical theory of scattering does not apply to short ranges potentials such as the `thorn' ones, 
%f the perturbing potentials 
$\delta\hat V$. It may overestimate the dechanneling rate by several ten percent. We have also pointed to a possible role of the correlated atom vibrations in dechanneling and radiation emission and suggested how to take them into account.    
 
\ni 
~ {
{\it Note}: the basic ideas of this work were first presented at the 8$^{\rm th}$ International Conference {\it Channeling 2018} (Sept. 23-28, 2018, Ischia, Italy). }
 
%{The most questionable approximation is to use the standard scattering theory, which normally applies to incident plane waves, while a channeling state is rather described by a wave packet of finite transverse size.} %  of a channeled wave function is finite.
%
%In the QM approach we assume that the range of $\delta V_i$ is much less than the transverse width of the wave function (local plane wave approximation). 
%
% A similar conclusion is reached for incoherent bremsstrahlung when calculated from the electron trajectories. 


\begin{thebibliography}{99}

\bibitem{Daba} S.B. Dabagov, V.V. Beloshitsky and M.A. Kumakhov,  {Nucl. Inst. Meth. in Phys. Research} B \textbf{74} (1993) 368; 
S.B. Dabagov and N.K. Zhevago,  Riv. Nuovo Cimento {\bf 31} (2008) 491.

\bibitem{Tikho19} V.V. Tikhomirov,  {\it Quantum features of high energy particle incoherent scattering in crystals}, Phys. Rev. Acc. Beams {\bf 22} (2019) 0504501.  

\bibitem{Kagan} Yu.M. Kagan and Yu.V. Kononets, Zh. Eksp. Teor. Fiz.  {\bf 64} (1973) 1042.

\bibitem{Kostyuk}  A. Kostyuk, A. Korol, A. Solov'yov, W. Greiner, {\it Planar channelling of 855 MeV electrons in silicon: Monte Carlo simulations}, J. of Phys. B {\bf 44} (2011) 075208. 
% (7) : Atomic, Molecular and Optical Physics 
% arXiv:1008.1707, doi:10.1088/0953-4075/44/7/075208.
% 3] A.Kostyuk,A.V.Korol,A.V.SolovÕyov,W.Greiner,Planarchanneling of electrons: Numerical analysis and theory;, Il Nuovo Cimento C 34 (4) (2011) 167Ð174. doi:10.1393/ncc/i2011-10931-9.

\bibitem{XAsimul} X. Artru, {\it A simulation code for channeling radiation by ultrarelativistic electrons and positrons},  {Nucl. Inst. Meth. in Phys. Research} B \textbf{48} (1990) 278.

\bibitem{Bondarenco} M.V. Bondarenco, {\it Atomic potentials and relationships between scattering observables}, Sept. 2019, to be published.

\bibitem{Bondarenco-rainbow} M.V. Bondarenco, private communication.

\bibitem{Glauber} R.J. Glauber, Phys. Rev. {\bf 98} (1955) 1692.

\bibitem{Jackson-cor} D.P. Jackson, B.M. Powell and G. Dolling, Phys. Lett. {\bf 51 A} (1975) 87.

\bibitem{Huan} Cheng Huan-Sheng, Chui Zhi-Xiang, Xu Hong-Jie, Yao Xiao-Wei and Yang Fu-Jia, {Nucl. Inst. Meth. in Phys. Research} B \textbf{45} (1990) 424.

\bibitem{XA-vibra} X. Artru, 
{\it Correlations in thermal vibrations of crystal atoms. Effect on dechanneling and bremsstrahlung}, 
{Nucl. Inst. Meth. in Phys. Research} B \textbf{402} (2017) 21.


\end{thebibliography}
\end{document}